\documentclass[aapt,ajp,reprint,amsmath,amssymb]{revtex4-2}
\usepackage{graphicx}
\usepackage[colorlinks=true,linkcolor=blue,citecolor=blue,urlcolor=blue]{hyperref}

\makeatletter
\g@addto@macro\UrlBreaks{\do\-\do\_}
\makeatother
\begin{document}

\title{A Tool-Invariant Framework for Teaching and Assessing Computational Methods in the Age of Agentic AI}

\author{Larry Engelhardt}
\email{lengelhardt@fmarion.edu}
\affiliation{Department of Physics and Engineering, Francis Marion University, Florence, SC 29505}

\date{\today}

\begin{abstract}
Learning a computational method has always meant learning to operate a tool---pencil, slide rule, calculator, or programming language. Agentic artificial intelligence, which writes, executes, and revises simulation code from natural-language specifications, is the latest and largest step in a centuries-long migration of mechanical work from human to tool. I argue that what a learner must know has remained remarkably stable: the inputs and outputs of a method, the concept of what it does, the terminology to communicate about it, the judgment to evaluate its results, and the skill of operating the current tool. This paper organizes these requirements into a tool-invariant framework spanning single-digit addition to agent-orchestrated molecular dynamics, argues that verification---not code authorship---is now the load-bearing skill, and draws the consequence for assessment: when artifacts can be generated on demand, the artifact no longer certifies the student. I describe a practical response, designed for the small classes where the subject lives---AI-free in-class coding quizzes paired with oral defenses of comment-stripped, AI-assisted work---and argue that the real product of a computational physics course is the student's ability to explain and defend computational artifacts in the language of the discipline.
\end{abstract}

\maketitle

\section{Introduction}
\label{sec:intro}

Consider what it means to know how to add. A child who computes $2+3=5$ knows more than a procedure: she knows what the inputs are (two positive whole numbers), what the output is (their combined total), that adding two positive numbers must yield something larger than either, and how to say all of this---\emph{plus} and \emph{equals}---to another person. Her first actuation tool was her own fingers, counted with her eyes---and the deepest of these ideas, that combining quantities makes the total grow, was in place before any tool at all. When the numbers grow to $53+71$, she needs an algorithm (line up the place values; carry) and a new tool: a pencil, and the fine-motor skill to write numerals with it. Later she will use a calculator, and the algorithm and the pencil skill will both become unnecessary---but everything else will not. She must still understand what addition is, still specify the inputs, still judge that 124 is a reasonable answer and that 12.4 is not, and still communicate about what she has done.

That pattern has repeated at every scale of scientific computation for more than two centuries, and it is repeating right now with unusual force. For the past decade, ``the tool'' meant a programming language and libraries---Python with NumPy and SciPy, for most of us---and ``operating the tool'' meant syntax, debugging, and the management of arrays and loops. Today the tool is increasingly an \emph{agentic} artificial intelligence: a system such as Claude Code that accepts a natural-language description of a physics problem, writes the simulation, runs it, plots the results, and revises its own work. The tool now writes the code.

Two things happened when it did, and this paper treats them as one event. First, the \emph{learning} question: if students no longer need to write the code, what exactly must they still learn? Second, the \emph{assessment} question: the traditional computational assignment---write code, generate plots, submit a report---collapsed as a measurement instrument, because the artifact a student submits no longer certifies anything about the student. I will argue that both questions have the same answer, and that the answer was visible all along in the arithmetic example above. In what follows I develop a framework of five tool-invariant requirements, apply it across a scaffolded progression of tasks (tabulated in full as supplementary material), and draw its consequences for teaching and assessment, ending with open questions our community needs to be discussing now. Parts of the paper are first-person and anecdotal by design: the field does not yet have controlled studies of students learning computational physics through agentic AI; it has practitioners adapting in real time, and I offer this framework as a structure for that adaptation---not as its conclusion.

This paper builds on four decades of computation in physics education---from M.U.P.P.E.T.\cite{redish1993} and \emph{Matter and Interactions}\cite{chabay2008} to the PICUP community,\cite{engelhardt_picup_book2025} the AAPT curricular recommendations,\cite{behringer2016,behringer2017} and Resource Letter CP-3\cite{atherton2023}---and on the computational-thinking tradition.\cite{wing2006,weintrop2016,weller2022,odden2019essays} Those frameworks catalog practices for students who \emph{write} code; this one re-weights them for students who \emph{direct} it. Published work on AI in physics education has so far centered on chatbots\cite{kortemeyer2023,polverini2024} and, most recently, on assessment---a survey of practice,\cite{sabo2026} a dedicated journal collection,\cite{prper2026} and an alarm over the collapse of unsupervised assessment;\cite{kortemeyer2026} computing education, roughly two years ahead on this problem, is pivoting from code writing toward specification and evaluation.\cite{denny2024cacm} The first empirical study of physics students working this way has just appeared: in interviews with nineteen undergraduates, the productive users limited the tool to small steps and verified constantly, while over-reliers accumulated false assumptions about their own models\cite{fredly2026}---early data that motivate this framework and warn against reading it as delegation cheerleading. To my knowledge, no published framework yet treats agentic AI as the primary \emph{mediating tool} for computational physics instruction. That is the gap addressed here, and Kortemeyer's assessment alarm\cite{kortemeyer2026} is the problem to which Sec.~\ref{sec:assessment} offers a constructive response.

Three elements here are offered as new: the principle of \emph{validation authority} and the bespoke-versus-socially-validated distinction that grounds it (Sec.~\ref{subsec:validation}); the tool-invariant \emph{re-weighting} argument, an organizing lens whose testable consequences are collected in Sec.~\ref{sec:openquestions}; and the two-instrument assessment design with its verification-gated oral defense (Sec.~\ref{sec:assessment} and supplement). The five pillars themselves largely repackage constructs the computational-thinking literature already contains (Table~\ref{tab:priorwork} maps the correspondence), re-weighted for students who direct code rather than write it.

\section{The long view: computation has always been delegated}
\label{sec:longview}

``Computer'' was a job title for two centuries before it was a machine;\cite{grier2005} a student directing an AI agent in 2026 is doing the job of the \emph{planners} who directed those human computers---decomposing the problem, designing the checks---and every student must now learn it. Each tool transition since has silently deleted a \emph{forcing function} for some habit of judgment: when the calculator deleted the slide rule's forced order-of-magnitude bookkeeping, educators feared for the estimation habit, and three decades of mathematics-education meta-analyses returned a verdict more instructive than the fear---calculator use does no harm, and often helps, \emph{precisely when instruction deliberately maintains the underlying number sense}.\cite{hembree1986,ellington2003} That conditional is the lesson on which this paper rests: \textbf{at each transition, the pillars of understanding remain, but some practice that used to train them incidentally stops happening by itself---so the compensation must be deliberate.} The larger the offloading, the more deliberate the compensation, and agentic AI is the largest offloading yet.

\section{The framework}
\label{sec:framework}

\subsection{Five pillars}
\label{subsec:pillars}

I claim that competent use of any computational method, at any scale, with any tool, requires five things:

\begin{enumerate}
\item \textbf{Inputs and outputs.} What goes in, in what form, under what conditions of validity; what comes out, and what it means. This includes posing the problem---deciding that a computation is worth doing at all.
\item \textbf{Conceptual understanding of the method.} Not the ability to execute the algorithm, but a working model of what it does---including its \emph{knobs} (step size, tolerance, sample count) and its \emph{characteristic failure modes} (instability, divergence, overfitting, aliasing).
\item \textbf{Terminology.} The vocabulary of the method and of the physics, precisely used.
\item \textbf{Sensemaking.} The ability to judge whether outputs make sense---in the physics-education-research sense of building an explanation that resolves a perceived gap or conflict in one's knowledge\cite{oddenruss2019}---and to \emph{establish} whether they are right (Sec.~\ref{sec:vv}). Computation reliably provokes exactly such conflicts, and that confrontation is where the learning lives.\cite{sand2019,kuo2020,sirnoorkar2023}
\item \textbf{Operating the tool.} The mechanical or interactional skill of actuating the computation: penmanship, keystrokes, syntax---or, now, directing an agent.
\end{enumerate}

The child adding $2+3$ exercises all five. So does the senior directing an AI agent to build a molecular dynamics simulation. What changes between those rungs is the \emph{content} of each pillar and the \emph{weight} a curriculum must give it---and the two move differently. Content changed most dramatically for pillar 5 (penmanship became prompting) and least of all for pillar 4; weight moved the other way: sensemaking, the pillar whose content barely changed, is the one the AI transition makes load-bearing (Sec.~\ref{subsec:validation}). Two pillars deserve immediate comment.

\textbf{Terminology is now an input channel.} When the tool accepts natural language, vocabulary becomes literally operative: ``integrate this ODE with an adaptive Runge--Kutta method and verify energy conservation'' and ``make the ball bounce right'' produce categorically different artifacts---not because the agent cannot handle the second (it might cheerfully guess), but because every choice that a vague specification leaves open is a choice the student has silently ceded to the tool. The precision of a student's disciplinary language is now written directly into the artifact, and it is hard to imagine a stronger argument for insisting on exact terminology.

\textbf{Tool operation has inverted its shape.} Operating NumPy well required narrow, deep skill. Operating an agent well requires broad, shallow fluency across many domains---programming ecosystems, file types, the command line, version control, the documentation that lets a project survive a cleared context, and the capabilities and costs of different AI model tiers. This is \emph{interactional expertise}:\cite{collins2007} fluency in a domain's language sufficient to converse, judge plausibility, and direct work, without the contributory expertise to do the work oneself. The AI-era computational physicist needs \textbf{contributory expertise in physics and interactional expertise in computing}; Sec.~\ref{subsec:toolfluency} offers a first-person illustration of what that combination can do.

\begin{table}[htb]
\caption{\label{tab:priorwork}Antecedents of the five pillars in three existing frameworks: Weintrop \emph{et al.}'s computational-thinking taxonomy for mathematics and science,\cite{weintrop2016} Weller \emph{et al.}'s computational-thinking practices for introductory physics,\cite{weller2022} and the computational-literacy construct (material/cognitive/social) adapted to physics by Odden \emph{et al.}\cite{odden2019essays} The pillars repackage this content; what is new is the re-weighting.}
\begin{ruledtabular}
\begin{tabular}{p{2.6cm} p{5.6cm}}
Pillar & Antecedent constructs \\
\hline
1. Inputs and outputs & Data practices and problem preparation;\cite{weintrop2016} specification-level practices\cite{weller2022} \\
2. Method concept & Modeling-and-simulation practices;\cite{weintrop2016} the cognitive dimension of computational literacy\cite{odden2019essays} \\
3. Terminology & The social dimension of computational literacy (disciplinary discourse)\cite{odden2019essays} \\
4. Sensemaking & Model-assessment and debugging practices;\cite{weintrop2016,weller2022} the sensemaking construct itself\cite{oddenruss2019} \\
5. Tool operation & The material dimension of computational literacy;\cite{odden2019essays} programming and tool practices\cite{weintrop2016,weller2022} \\
\hline
What is new & The re-weighting under agentic mediation; validation authority (Sec.~\ref{subsec:validation}); the assessment consequence (Sec.~\ref{sec:assessment}) \\
\end{tabular}
\end{ruledtabular}
\end{table}

\subsection{The pillars in action: iterating}
\label{subsec:loop}

The pillars catalog five essential ingredients for effective use of computation, and in practice they organize into a workflow,

\begin{center}
\textbf{Specify $\to$ Predict $\to$ Delegate $\to$ Verify $\to$ Interpret; Iterate}.
\end{center}

\emph{Specify} the problem, its inputs, outputs, and physical model (pillar 1, resting on pillars 2--3). \emph{Predict} the result before any execution:\cite{white1992,sentance2019} order of magnitude, sign, units, limiting behavior---a Fermi estimate of the answer.\cite{mahajan2010} A student who has committed to an expectation is \emph{testing a hypothesis} when the plot appears; a student who has not is merely \emph{looking at a picture}. \emph{Delegate} the execution to the tool (pillar 5). \emph{Verify} the output against the prediction and the checks of Sec.~\ref{sec:vv} (pillar 4). \emph{Interpret} the result physically, reconcile any discrepancy, and iterate.

One more competency threads through the loop, trivial for deterministic tools and essential now: \textbf{calibrated reliance}\cite{lee2004}---knowing, for this tool on this task, how much verification is owed. A calculator's arithmetic needs a glance. A community-validated library routine needs interface-level checks. A bespoke, agent-generated simulation needs the full battery, every time. The next subsection says why.

\subsection{Why verification is newly load-bearing}
\label{subsec:validation}

An obvious objection: physicists have trusted black boxes for decades. Nobody reads the source of a library FFT or linear-algebra routine before calling it, and nobody should. If opacity were the problem, this framework would have been urgent in 1995.

But opacity was never the problem;\cite{humphreys2004,humphreys2009} \emph{validation} is. A library routine is opaque but \textbf{socially validated}---published algorithms, decades of testing, millions of users whose combined experience would have surfaced any systematic error. Trust in it is rationally grounded in the reliability of the process that produced it.\cite{duran2018} An AI-generated simulation is opaque and \textbf{bespoke}: an artifact with a population of one, produced by a stochastic process whose competence boundary is jagged and invisible,\cite{dellacqua2026} and whose failures arrive disguised as successes---running code, smooth plots, confident prose. A wrong trajectory renders as beautifully as a right one.

\textbf{The verification burden that library ecosystems amortized across the entire community now lands on each student, for each artifact, every time.} That sentence is why the sensemaking pillar, always present, becomes load-bearing in the AI era in a way it never was for SciPy.

A corollary---call it the principle of \textbf{validation authority}---generalizes the point: \emph{delegation is safe exactly where the delegator retains the ability to validate the outputs.} This says what a computational physics course is now \emph{for}: it trains validation authority over physical simulations. It also marks the boundary of the observation that ``anyone can build software with AI now.'' Anyone can---in domains like personal productivity apps, where correctness is \emph{observable in use}. A physics simulation's correctness is not observable in use; it must be \emph{established}, by disciplinary checks the user has to know to demand. The plots do not debug themselves.

\subsection{What must remain human}
\label{subsec:human}

A framework whose slogan is ``delegate the mechanics'' owes readers the non-delegable list:

\begin{enumerate}
\item \textbf{Posing the problem.} Regardless of whether or not the agent asks good clarifying questions---today's agents routinely do---deciding what is worth computing is the scientist's responsibility.
\item \textbf{Choosing the physical model and owning its assumptions.} The agent will implement a wrong model beautifully.
\item \textbf{The prediction}, made before execution, without the tool.
\item \textbf{Specifying the checks.} Deciding what counts as evidence of correctness \emph{is} the scientific judgment; delegating it to the system being checked is circular.
\item \textbf{Final epistemic responsibility.} ``The AI said so'' is not a justification available to a scientist.
\end{enumerate}

Items 1, 2, and 5 are not claims about current AI capability, and do not weaken as models improve. They are constitutive of doing science.

A full task-by-task progression applying the framework---from single-digit addition to agent-orchestrated molecular dynamics, recording at each rung what the era's tool absorbs and what must remain human---is given as supplementary material (Table~S1). A companion table there (Table~S2) holds one task fixed and walks it across five tool eras, isolating the framework's central claim: only the actuation skill and the division of labor change; the Specify, Predict, and Verify work never does.

\section{Sensemaking, operationalized: verification and validation for the classroom}
\label{sec:vv}

``Check whether the answer makes sense'' is an exhortation, not a curriculum. Professional computational science long ago turned it into one: \textbf{verification} (are we solving the equations right?), \textbf{validation} (are we solving the right equations?), and uncertainty quantification.\cite{oberkampf2010} Scaled to the classroom, the sensemaking pillar becomes a specific, cumulative repertoire. Each check enters at a particular rung of the progression (supplementary material, Table~S1) and remains required at every rung thereafter.

\emph{Verification-type checks:}
\begin{itemize}
\item \textbf{V1 --- Units and dimensions.} Introduced with word problems; universal thereafter.
\item \textbf{V2 --- Order of magnitude and bounds.} The Fermi estimate.
\item \textbf{V3 --- Known-answer check.} Run the method on a case with an exact solution before the case without one: integrate $x^2$ before $e^{-x^2}$; simulate the drag-free projectile before adding drag.
\item \textbf{V4 --- Self-consistency.} Multiply back, substitute back, differentiate the integral, Parseval for the FFT.
\item \textbf{V5 --- Convergence.} Halve $dt$, double $N$; the answer should stabilize, and how fast is itself diagnostic.
\item \textbf{V6 --- Conservation.} Energy, momentum, probability, particle number. The workhorse for all simulation.
\item \textbf{V7 --- Independence.} Three practices shelter under this name, and they are not equally strong. (a) Solving the problem by a genuinely different method is the gold standard. (b) Rerunning stochastic code with a fresh seed probes stochastic stability---for Monte Carlo, the standard error of the mean falls as $1/\sqrt{N}$. (c) Re-querying the \emph{agent} is the weakest: two runs of the same model are correlated samples from one trained distribution, sharing its systematic biases, so their agreement is weak evidence;\cite{knight1986} asking twice is a crude probe of the tool's variance, not a second opinion---a lesson in correlated errors that outlives any particular tool.
\end{itemize}

\emph{Validation-type checks:}
\begin{itemize}
\item \textbf{V8 --- Limiting cases.} Turn off drag and recover the parabola; let $T\to 0$ and recover the ground state.
\item \textbf{V9 --- Comparison with experiment or data}, where available.
\end{itemize}

Two properties deserve emphasis. First, \textbf{every check in this repertoire is physics or mathematics content, not programming content}: a sensemaking curriculum fits in a physics course without teaching software engineering. Second, the checks are \textbf{cumulative and cheap}---most cost minutes---which is what makes them enforceable as culture rather than aspirational as policy. A third property must be stated with equal candor: these checks \emph{bound} the space of undetected errors; they do not close it. A stronger tier---\emph{code verification} by manufactured solutions and order-of-accuracy tests\cite{oberkampf2010}---requires engaging the code as code, contributory expertise this course deliberately does not build. And in the capstone workflow the agent typically executes the checks, so each check inherits the trust problem of the system under test; the governing rule is item 5 of Sec.~\ref{subsec:human} applied without exception: the system being checked is never the sole authority on whether it passed.

\section{Teaching practices}
\label{sec:practices}

\subsection{White-box, then black-box}
\label{subsec:whitebox}

Buchberger's white-box/black-box principle, from computer-algebra education, is this paper's sequencing rule: a method should be studied transparently before it is invoked opaquely, and yesterday's white box becomes tomorrow's trusted primitive.\cite{buchberger1990} In my courses each method class gets a transparent white-box phase before delegation: students build the fifteen-line Euler integrator, watch it fail at large $dt$, and must implement it on an AI-free quiz before the course moves to \texttt{solve\_ivp} and, later, to directing agents. The white-box phase is sized for conceptual residue---the working model of pillar 2---not for professional fluency. This one rule dissolves most of the apparent conflict between ``students must code'' and ``students need not code'': both are true, at different rungs, for different durations. What the rule does not settle is the dose---whether or not a brief white-box phase suffices is the framework's central untested assumption (Sec.~\ref{sec:openquestions}, question 1).

\subsection{Error injection and specification exercises}
\label{subsec:errorinjection}

Students receive a subtly wrong simulation\cite{cox2011}---sign error in the force, $dt$ an order too large, the wrong potential---and the graded object is the diagnosis. Several of my in-class quizzes have taken exactly this form for years, though the richest source of injected errors is now real agent sessions. The exercise provides the supervisory practice that the agentic workflow no longer provides by itself,\cite{bainbridge1983} and it is graded practice in \emph{reading code one did not write}---the fluency that the defense's walkthrough will demand (Sec.~\ref{subsec:defense}), trained here and by the quizzes' provided-code format rather than by writing at home. Its complement, computing education's ``prompt problems,''\cite{denny2024prompt} runs the other direction: given a target behavior, author the specification that reliably produces it.

\subsection{What tool fluency looks like: a first-person account}
\label{subsec:toolfluency}

I resisted agentic tools until a week-long, in-person PICUP workshop in June 2026. Within a week of returning I had built an offline dashboard for my institution's learning-management system\cite{engelhardt_bbdash}---including the comment-stripping tool on which Sec.~\ref{sec:assessment} depends---without writing the code myself. That week demanded knowing a little about a lot, including the language and concepts to communicate precisely with the system doing the work. Note the closed loop: the same class of tools that broke my old assessment scheme \emph{built the infrastructure of its replacement}, in days, by one professor.

Two caveats keep this account honest, and both cut in the framework's favor. The one-week curve was possible \emph{because} the conceptual pillars were already deep; only the surface was missing, and the surface is now cheap---for our students, building the pillars is what the course is for. And a dashboard is the \emph{easy} case by this paper's own distinction (Sec.~\ref{subsec:validation}): its correctness is observable in use. A physics simulation's is not---first-person evidence from the easy case cannot carry the framework's hard claim. That hard case, verification where correctness must be \emph{established} rather than observed, is what the repertoire of Sec.~\ref{sec:vv} exists to train; and I will be testing with students this fall, including assignments in quantum mechanics.\cite{engelhardt_picup_qm} 
Additionally, a complementary PICUP exercise directing a large language model through the double pendulum has recently been prepared.\cite{colleague_picup_dp}

\subsection{Motivation, and the limits of guardrails}
\label{subsec:motivation}

In Fall 2025 I built a customized chatbot for my sophomore-level computational course (about fifteen students)---a tutor with guardrails, designed for Socratic dialogue rather than answers. I have no systematic data---an end-of-semester survey drew exactly one response, itself a small lesson in instrumentation---so what follows is one instructor's reading of one semester: the response split sharply. For students willing to engage, it was excellent---consistent with early randomized-trial evidence (a single two-week study in one introductory course) that a well-designed AI tutor can outperform in-class active learning;\cite{kestin2025} for students who wanted the answer, it was useless, because an unguardrailed model is two browser tabs away. AI removes nearly all friction from \emph{executive} help-seeking---wanting the answer rather than the understanding.\cite{nelsonlegall1981} The structural lesson: \textbf{outside of class, ``you may not use AI'' is not a policy; it is a wish.} The framing I now offer students is borrowed from the writer Ted Chiang:\cite{chiang2024} homework is the gym, not the job. Nobody lifts weights because society needs the weights moved; the lifting builds the capacity for the things society does need. An AI that can do your homework is a forklift at the gym---stronger than you will ever be, and beside the point. The design problem is therefore not prohibition but motivation---course structures in which doing the work is the rational strategy, and in which competence, autonomy, and relatedness\cite{ryandeci2000} have room to operate. The assessment design of the next section is my answer: competence regains meaning (the defense cannot be faked), autonomy grows (AI-assisted production lets students attempt more ambitious simulations than their coding fluency alone would allow), and the course culminates in the highest-relatedness assessment format there is---a one-on-one conversation with a physicist about work the student owns.

\section{Assessment: what is the product?}
\label{sec:assessment}

\subsection{The proxy collapse}
\label{subsec:proxy}

For two decades I assigned computational work in the traditional mode: write code, run it, generate plots, submit a report. That mode is dead, and it is worth being precise about what killed it and what, exactly, died. The report was never the thing we cared about; it was a \emph{proxy} for understanding, valid only because producing the artifact required the understanding. Goodhart's law---``when a measure becomes a target, it ceases to be a good measure''\cite{strathern1997}---was held at bay exactly as long as gaming the measure cost as much as learning. AI dropped that cost to zero, and the proxy collapsed:\cite{kortemeyer2026} performance \emph{with} the tool no longer provides evidence of a student's capability \emph{without} it.\cite{salomon1991} What died is specifically the \emph{unsupervised artifact}. Supervised formats---proctored practical exams, process-based grading over version histories---survive, and one of them is half of the design below. AI did not break assessment; it revealed that assessment was always measuring an artifact and inferring a student---and the inference no longer holds.

So what \emph{is} the product of a computational physics course? Not the code, the plots, or the report: the world now has an essentially unlimited supply of those at negligible marginal cost. \textbf{The product is the student's ability to explain and defend the artifacts---code, plots, claims---using the concepts and language of the discipline.} A student who can do that can direct computational work responsibly for a career, whatever the tools become. The academy has already solved this problem once, at its apex: the Ph.D.\ is certified not by the dissertation artifact alone but by its \emph{defense}, precisely because dissertations are produced collaboratively and the artifact alone cannot certify the candidate. AI has now made every student's artifact collaborative. Nor is the instrument my invention: oral assessment has a developed research literature,\cite{joughin1998,iannone2015} and computing education has built and evaluated code interviews in direct response to generative AI.\cite{kannam2025} What follows adapts that instrument to computational physics, with one addition I regard as the physics-specific core: the verification gate.

\subsection{Two instruments}
\label{subsec:twoinstruments}

My own course design has passed through overlapping stages rather than a clean sequence: two decades of artifact grading; specification-based pass/not-pass submissions since Fall 2019 (a pass requires meeting all ten published specifications at A/B quality; resubmission allowed at a modest penalty); weekly AI-free coding quizzes in a lockdown browser since Fall 2023; the guardrailed tutor of Fall 2025 (Sec.~\ref{subsec:motivation}), which taught me that tool-side guardrails cannot substitute for assessment-side design; and---beginning that same semester---ten-minute oral defenses following each assignment, in Classical Mechanics and Advanced Computational Physics (both Fall 2025). Fall 2026 commits the sophomore computational course to the full design: AI-permitted project work, assessed by the formalized protocol of Sec.~\ref{subsec:defense}. Notice what the AI transition did to the oldest instrument: the report specifications---complete, well written, readable code, correct results, discussed plots---are precisely what agents now supply by default. The specifications did not become wrong; they became free---so I have retired specification grading, since an artifact a tool satisfies on demand no longer certifies its author. The proxy collapse of Sec.~\ref{subsec:proxy}, enacted in my own course.

What is tested, and what is not. The quizzes are three fall semesters of practice at roughly fifteen students each; some are error-injection exercises outright (broken code to debug); they run in a browser-embedded Python environment\cite{trinket} where students receive correct import statements and often a minimally working template---the tedious scaffolding---and implement the concept being assessed, with access to documentation through \texttt{help()} along the way. Oral examinations have been standard in my upper-level teaching since 2018---six offerings of a computational statistical-and-thermal-physics course, whose oral exams probe computational understanding directly. The per-assignment oral defenses are newer: one semester, in two upper-level courses with enrollments of three and five---without comment-stripping and without a formal rubric. Together that experience taught me two things: ten minutes suffices to tell whether a student can connect the concepts, the terminology, and their own artifact; and the strongest students treat the session as a stage rather than a trial. What is genuinely new this fall, and therefore genuinely untested: the automated comment-stripping, the published rubric with its verification gate, and the sophomore-level population. This paper, then, reports experience where it exists and design where it does not; the first full-protocol data arrive with the Fall 2026 cohort.

The quizzes and the defenses are not competing strategies, and neither is transitional---they are the two instruments the framework demands. The quizzes assess the \textbf{white-box phase}: if coding fluency is a learning goal, AI-free conditions are now the only conditions under which it can be validly measured. The defenses assess \textbf{orchestration}: the full loop of Sec.~\ref{subsec:loop}, exercised on AI-assisted work. A course with only quizzes certifies residue but never orchestration; a course with only defenses certifies orchestration built on unverified foundations.

\subsection{The oral defense protocol}
\label{subsec:defense}

The protocol: students submit their code and plots; AI use is permitted and disclosed. Before the defense I strip all comments from the submitted code with an automated tool. We then meet one-on-one for a maximum of ten minutes---a semester of the informal predecessor, atop years of oral examinations, established that ten is enough (Sec.~\ref{subsec:twoinstruments}): a brief uninterrupted overview by the student; a walkthrough of two or three instructor-chosen regions of the uncommented code---the student's \emph{own} project, iterated on for a week, so the demand is familiarity with one's own artifact, not cold review of foreign code; an explanation of each plot in physical terms; and a round of verification probes---\emph{why should I believe this?}

Why strip the comments? AI-generated code arrives \emph{pre-narrated}: fluently commented, so that understanding can be performed by reading the narration aloud. Stripping deletes the cheapest performance---in the room, the explanation must come from somewhere other than the screen. But stripping is hygiene, not the load-bearing wall, and it is worth being adversarial about why: a student can paste their own stripped code back into an agent the night before and rehearse fluent, purpose-level answers. The narration is deleted from the file, not from the student's reach. What actually certifies is the part that cannot be rehearsed in advance, because it does not exist in advance: the probing is \emph{live and adaptive}---the follow-up depends on the student's previous answer; the counterfactual (``if I doubled the mass, which plots change?''); the request to point to the line where Newton's second law enters the program. And the deeper reply is constructive alignment itself:\cite{biggs1996} a student who drills with an AI until they can survive adaptive probing of their own code has, in the drilling, built exactly the understanding the course intends---explaining is not merely evidence of understanding but a producer of it.\cite{chi1994} The workaround is the work. A semester of these defenses---run before I had the stripping tool---taught me that genuine and performed command separate within minutes under live probing: purpose (``this is the Euler velocity update---it's where Newton's second law enters'') versus syntax paraphrase (``this sets \texttt{v} equal to \texttt{v} plus \texttt{a} times \texttt{dt}''). What stripping adds---deleting the borrowed narration before the student sits down---meets its first full test this fall.

The defense is scored on a rubric of five dimensions aligned with the framework---code comprehension, method understanding, physics model and terminology, interpretation of results, and verification; every pillar is assessed---each on a four-level scale, with a gate: the verification dimension must reach the ``functional'' level for the defense to pass, regardless of total. (The full rubric, with leveled descriptors, a probe bank, and logistics, is provided as supplementary material; its thresholds are one examiner's working defaults, informed by a semester of informal predecessors and years of oral-examination experience, and awaiting the calibration that only full cohorts can supply.) One standard probe deserves mention: \emph{``What did the AI decide that you didn't? Did you override anything?''}---calibrated reliance, assessed directly, and honest AI use normalized rather than driven underground. The rubric and sample probes are published to students in advance, deliberately: by constructive alignment, students study what the assessment demands, and here it demands exactly the learning we want. A student who knows the defense is coming cannot let comprehension debt\cite{comprehensiondebt} accumulate during the week; the debt comes due, orally, with interest. Two notes connect the defense to the rest of the course. The submission includes a five-line log (the prediction, plus which checks of Sec.~\ref{sec:vv} were run, with outcomes), which the defense probes. And the defense is scored once, on a traditional 0--100 scale, with no retakes; the course's mastery-oriented machinery---pass/not-pass with penalized retries---lives in its weekly quizzes instead (Sec.~\ref{subsec:twoinstruments}).

\subsection{Scalability---and why ``it doesn't scale'' is partly the point}
\label{subsec:scalability}

Honesty requires the arithmetic. The ten-minute session length is measured---a semester of per-assignment defenses in two courses stands behind it; the fifteen-student cohort is this fall's projected enrollment, so the product is an estimate: ten minutes times fifteen students is two and a half contact-hours per defended assignment---\emph{less} than conscientious grading of fifteen reports, and incomparably more informative. Two things temper that figure: at roughly eleven defended assignments per term, the course projects to some twenty-eight contact-hours of defenses per semester; and these are additive to the weekly AI-free quizzes, which measure a different thing (Sec.~\ref{subsec:twoinstruments}). My summary after one semester of the practice, at smaller enrollments: a good investment of instructor time. For the small classes where this subject lives, the trade is available. For the three-hundred-seat course it plainly is not, and that reader is owed more than a slogan. Three degraded modes exist, each with a named cost: five-minute spot-defenses on two rotating, unannounced dimensions \emph{sample} understanding rather than certify it, but preserve the incentive, because no student knows in advance which dimensions are theirs; calibrated teaching assistants scale the format---computing education has run TA-led code interviews in large courses\cite{kannam2025}---at the price of examiner variance that rubric descriptors and shared scoring only partly contain; paired defenses, students probing students, change the instrument entirely and belong in formative rehearsal, not certification. And one consequence should be said aloud: if trustworthy certification requires human-scale assessment, then students at institutions that cannot staff it receive systematically less trustworthy credentials---an equity problem, and an argument faculty should be making to their administrations. The oral defense does not scale to three hundred students---\emph{and that is information about the three hundred, not about the defense}: class size has become an academic-integrity variable, and small classes have acquired a new and concrete justification.

\section{Objections and replies}
\label{sec:objections}

\textbf{Deskilling---the ironies of automation.} Supervisory judgment is built by the routine practice that automation removes,\cite{bainbridge1983} and supervisory skills measurably decay;\cite{casner2014} students who never wrestle with the integrator will lack the mental model needed to catch the agent's errors. \emph{Reply:} This is the framework's motivating worry, and the design addresses it structurally---it does not yet answer it empirically. The white-box encounters (Sec.~\ref{subsec:whitebox}) and error-injection exercises (Sec.~\ref{subsec:errorinjection}) \emph{manufacture deliberately} the practice the workflow no longer provides incidentally, and because the quizzes recur weekly, the residue is re-exercised all semester rather than certified once and left to decay. Whether the dose suffices is open question 1: the pilots' manual-flying skills decayed under training doses far larger than a brief white-box phase.\cite{casner2014} The design is a bet, stated so that it can be tested.

\textbf{Constructionism---building is where the learning lives.} The most successful computation-in-physics curricula are premised on students constructing models,\cite{chabay2008,odden2019essays} and mathematics-education research warns that technique and concept are dialectically linked: removing technique does not automatically liberate concept.\cite{artigue2002} \emph{Reply:} The framework rations construction; it does not eliminate it (row 9 of the supplementary progression---the hand-coded ODE integrator---is mandatory and transparent). What it declines to require is \emph{comprehensive} construction---and the community's own practice is the precedent: nobody requires students to build an FFT, and nobody calls that a constructionist betrayal. The question was never whether to black-box, but when.\cite{buchberger1990} One concession the precedent does not cover: the ODE loop is not the FFT---it is where Newton's second law becomes a differential equation before a student's eyes, and the confrontations that drive computational sensemaking arise largely \emph{during} construction.\cite{sand2019,kuo2020} That is why row 9 is the one mandatory build, and why ``one encounter'' may be too few. The dose dispute is not resolved here; it is question 1, and the constructionist tradition may yet win it.

\textbf{``The AI will soon be reliable enough that checking is theater.''} \emph{Reply:} Three of the five non-delegables (Sec.~\ref{subsec:human})---posing the problem, owning the model, bearing epistemic responsibility---are not reliability questions at any capability level. As for the rest: reliability is a track record an artifact \emph{earns}, and a bespoke artifact with a population of one has none until its user supplies it; and ``reliable at what?'' is exactly the jagged-frontier problem.\cite{dellacqua2026}

\textbf{Access and equity.} Frontier agentic tools cost money. \emph{Reply:} Every rung of the supplementary progression (Table~S1) through row 11 runs on free tools, and the verification curriculum costs nothing---but the concern is real at the capstone, institutional licensing matters, and the community should address it collectively.

\section{Open questions for the community}
\label{sec:openquestions}

In 2026 these are genuinely open questions; the answers they eventually get will themselves evolve. This is the list I most want my colleagues arguing about.

\begin{enumerate}
\item \textbf{How much white-box is enough---and does the verification curriculum actually calibrate students?} These are the framework's central untested assumptions, and both are testable in our courses now: Does the hand-coded Euler encounter improve later diagnosis of black-box failures? Does prediction-first measurably improve detection of injected errors, or do students learn to perform checking as ritual? The instrument needs the same scrutiny: the rubric and its gate are one examiner's unvalidated defaults; inter-rater reliability, gate-failure rates, and differential impacts by anxiety and language background are measurements the first cohorts should supply.\cite{joughin1998,iannone2015}
\item \textbf{What is the fate of coding fluency as a learning goal?} I hold the ``permanent but small'' position---some direct coding, honestly assessed in AI-free conditions, indefinitely. Colleagues I respect hold both neighboring positions; the question is empirical, not settled. Its sharper form: how much code \emph{reading} does directing agents require, and can reading be trained without proportional writing? The defense's walkthrough dimension presumes an answer; my courses bet on error-injection exercises and provided-code quizzes as the training ground.
\item \textbf{How do we build intrinsic motivation at scale?} The defense model works partly because it is small and human. What preserves competence, autonomy, and relatedness in the 300-student service course?
\item \textbf{How do we develop faculty?} My own conversion required a week of in-person immersion among practicing colleagues---as the expertise literature predicts.\cite{collins2007} PICUP's workshops are infrastructure built for exactly this; what is the plan for scaling it to the majority of physics faculty who have not yet worked alongside an agent?
\item \textbf{What is a computational physics \emph{course} in 2036?} Does it dissolve into every course---computation as a medium---or sharpen into the discipline of specification and verification described here? The half-century of paradigm stability behind us is over; curricular structures should be designed to be revised.
\end{enumerate}

\section{Conclusion}
\label{sec:conclusion}

What the AI transition changes is the weighting: actuation is cheap now; judgment is not---and because each artifact an agent produces is bespoke, validated by no community and tested by no ecosystem, verification has become the discipline's teachable core while artifact-based assessment collapses.

The collapse is clarifying. The product of a computational physics course was never the code or the report; those were proxies, honest only while they were expensive. The product is the student who can stand next to a result and defend it---explain what was computed and why, what was assumed, what was checked, and why the claim deserves belief---in the precise language of physics. That ability is certified the way the academy has always certified it at its highest level: in person, in dialogue, one physicist to another. The artifacts got cheap. The conversation did not, and there is a deep sense in which the conversation is now the course. This is the part that we humans will do---and it adds value that a human is doing it.

\begin{acknowledgments}
This paper grew directly out of a week-long, in-person PICUP workshop in June 2026; the interactions there are what led to this paper, and I thank my PICUP colleagues who attended. The author is a principal investigator on the NSF-funded PICUP project, and this material is based in part upon work supported by the National Science Foundation under Grant No.~2337050; any opinions, findings, and conclusions expressed are those of the author and do not necessarily reflect the views of the NSF.

\emph{Disclosure of AI use.} This paper was prepared in collaboration with generative AI tools---Google's Gemini 3.1 Pro and Anthropic's Claude (Opus 4.8 and Fable 5)---used in the manner the paper itself describes. I specified the thesis, the framework, the structure, and ideas from my own teaching; I delegated literature search, drafting, and citation work to the AI systems; I verified all AI generated results; and I iterated through this process several times to refine the result. Responsibility for every claim remains mine alone---which is, of course, the paper's fifth non-delegable in action.
\end{acknowledgments}

\end{document}